\documentclass[a4paper,showkeys,floatfix,aps,pre,longbibliography,superscriptaddress,reprint]{revtex4-1}
\usepackage{graphics,graphicx}
\usepackage{amsmath,amssymb}
\DeclareMathOperator{\sech}{sech}
\usepackage{graphics,graphicx}
\usepackage{dcolumn,bm}
\usepackage{psfrag}
\usepackage{xstring}
\usepackage{color}
\usepackage{booktabs}
\newcommand{\srm}
{\affiliation{Department of Physics, SRM University - AP, Amaravati,
 Andhra Pradesh - 522240, India}}

\begin{document}

\title{Universal critical phase diagram using Gini index}

\author{Soumyaditya Das}

\email{soumyaditya\_das@srmap.edu.in}
\srm
\author{Soumyajyoti Biswas}

\email{soumyajyoti.b@srmap.edu.in}
\srm
\begin{abstract}
The critical phase surface of a system, in general, can depend on one  or more parameters. We show that by calculating the Gini index ($g$) of any suitably defined response function of a system, the critical phase surface can always be reduced to that of a single parameter, starting from $g=0$ and terminating at $g=g_f$, where $g_f$ is a universal number for a chosen response function in a given universality class. We demonstrate the construction with analytical and numerical calculations of mean field transverse field Ising model and site diluted Ising model on the Bethe lattice, respectively. Both models have two parameter critical phase surfaces -- transverse field and temperature for the first case and site dilution and temperature in the second case. Both can be reduced to single parameter transition points in terms of the Gini index. We have additionally demonstrated the validity of the method for a mean field two parameter opinion dynamics model that includes a tri-critical point.  The method is generally applicable for any multi-parameter critical transition.    
\end{abstract}

\maketitle

\section{Introduction}
A system can reach a critical point by fine tuning its external parameter(s). In many cases, there can be multiple such parameters. For example, while in the pure Ising magnets the external temperature is the sole tuning field through which the system can be brought to the Curie point, in site diluted version of the model, even at a constant temperature one can change the occupied site concentration to reach the critical point \cite{berger}. Another such example is the transverse field Ising model \cite{sach,tim}, which has transverse field and temperature as the tuning fields, and the critical phase surface can be crossed by tuning either or both of these parameters. Other such examples exist in a wide variety of situations from liquid crystals \cite{liq} to ecological transitions \cite{pnas}. 

In general, for a system with $r$ tuning fields, the critical phase surface could be up to $r-1$ dimensional hyperplane. Without analytical estimates, that exist only in rare instances, numerical/experimental determination of such a phase surface could become challenging through individual tuning of each of the tuning fields. For example, in cases of studying transition in vegetation patterns, the tuning fields (dependent on environment) cannot be varied in a controlled manner; in the case of fracture of porous rocks, the porosity (can be thought of as site dilution) is often indirectly estimated, but the critical point (critical load) is a sensitive function of it.  However, there are cases of special interests where an accurate estimate of the proximity to the transition point is of vital importance. Such situations include systems with potentially catastrophic transitions, e.g., transitions in plantation yields \cite{pnas}, ice coverage in polar regions \cite{polar}, breakdown of disordered solids \cite{wiley_book} to name a few. 

In this work, we provide a framework that allows for a simplification of a potentially $r-1$ dimensional critical phase manifold to a single parameter phase diagram. The parameter in question is a suitably defined Gini index for a diverging response function of the system, and the resulting critical point value is solely a function of the (universal) critical exponent value of the chosen response function. This means, so long as the universality class of the model is known, an estimate of imminent criticality (and critical scaling behavior) can be made independent of the details of the system that may often be inaccessible.   We start with a Landau theory formulation for a general $r$ parameter model and calculate the Gini index for diverging susceptibility and thereby arrive at a simplified phase diagram in terms of the Gini index of that diverging response function. We then demonstrate the applicability of the method through mean field calculations for the transverse field Ising model, a site diluted Ising model on the Bethe lattice and a mean field opinion dynamics model having also a tri-critical point.  

  \begin{figure*}
\includegraphics[width=18cm]{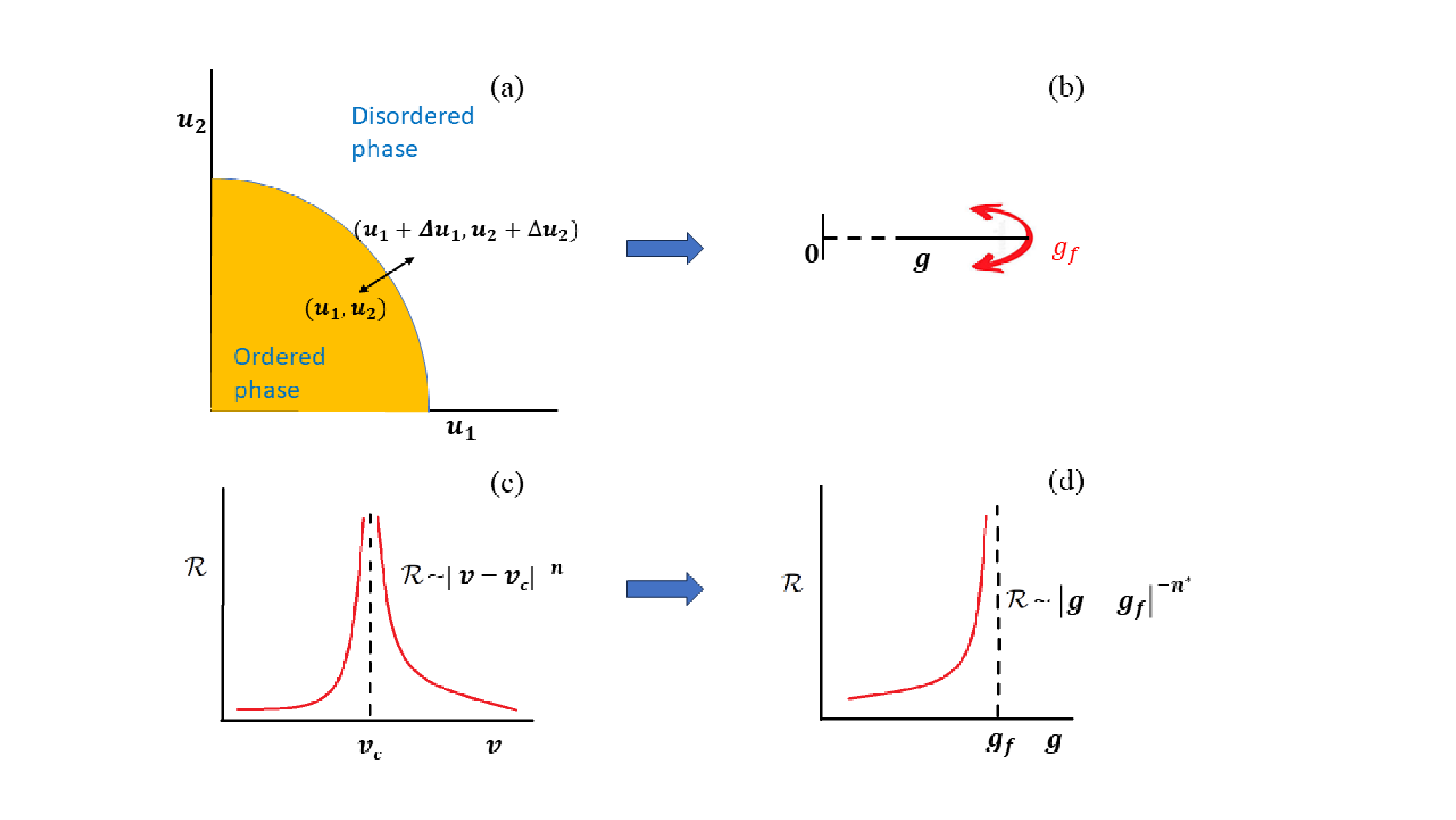}
\caption{A schematic diagram is shown to illustrate how a generic phase diagram of two parameters could be replaced by a line starting from $g=0$, terminating at a universal number $g=g_f$. $(a)$  A schematic phase diagram of two parameters, $u_1$ and $u_2$. $(b)$ Now the phase diagram of two parameters can be reduced to a line for some response $\mathcal{R}$ in $(c)$, when written in terms of their inequality in  $(d)$ close to the critical point .}
\label{fig_0}
\end{figure*}

\section{Formalism} The Landau free energy $G$ in the vicinity of a critical point can be expanded in a Taylor series of the order parameter $\phi$ (see e.g., \cite{berger}) up to the minimal relevant terms, using the analytic nature of $G$ in $\phi$ and the symmetry $G(\phi)=G(-\phi)$ as follows:
\begin{eqnarray}
    G([u_i],\phi)&=&a_0(u_1,u_2,....,u_r) +a_1(u_1,u_2,....,u_r)\phi^2  \nonumber \\
    &+&a_2(u_1,u_2,....,u_r)\phi^4-h\phi;
\end{eqnarray}
where the coefficients have their usual meanings, and $h$ is an external field that couples linearly with the order parameter $\phi$. We get the equilibrium order parameter by minimizing the free energy. The equation of phase hyperplane, as mentioned before, is $a_1(u_1,u_2,....,u_r)=0$  and this hyperplane could be crossed along any path say $v=v(u_1,u_2,....,u_r)$. Now, an infinitesimal change along the path $\Delta v=\sum_{i}\frac{\partial v}{\partial u_{i}}\Delta u_{i}$ (up to first order) relates to the order parameter as $\phi \propto {\Delta v}^{1/2}$. Hence the order parameter can be written as 
\begin{equation}
    \phi \propto {\Delta v}^{1/2} \propto\left[\sum_{i}\frac{\partial v}{\partial u_{i}}\Delta u_{i}\right]^{\frac{1}{2}}
    \label{L_op}
\end{equation}

Similarly for a susceptibility function one has,
\begin{equation}
     \chi \propto {\Delta v}^{-1} \propto{\left[\sum_{i}\frac{\partial v}{\partial u_{i}}\Delta u_{i}\right]}^{-1}.
     \label{L_chi}
\end{equation}

 As an example, for a transverse Ising model, there are two tuning fields viz. temperature ($u_1=T$) and transverse field ($u_2=\Gamma$). Following the discussions above, one would expect that the susceptibility $\chi \propto (A_1\Delta T+ A_2\Delta \Gamma)^{-\gamma}$, where $A_1$ and $A_2$ are at least analytic functions of $T$ and $\Gamma$. We shall soon see that it is indeed the functional form in this case. 
Also, even though the above formulation is for a mean field scenario, it is reasonable to express any diverging response function for any system characterized by multiple driving fields as $\mathcal{R}\propto {\Delta v}^{-n}$, where $n$ is the corresponding critical exponent \cite{polar}.

A diverging response function (that can be susceptibility or can be something else, as wel shall see later) would register highly unequal values. This is of course the result of critical fluctuations in the vicinity of critical transitions, irrespective of how the transition is approached (it is often hard to fine tune one or the other parameter in a controlled manner, see e.g. \cite{pnas}). It has been shown recently that a quantification of the inequality for the observed values of a diverging response function can lead to a major simplification of the scaling properties of such a function \cite{das}. Specifically, the inequality of the response function can be quantified by measuring the Gini index, which is traditionally used for quantification of socio-economic inequality \cite{gini}. It is usually defined through the Lorenz function $\mathcal{L}(f)$, which gives that $f$ fraction of the smallest values represent $\mathcal{L}(f)$ fraction of the total values. In terms of wealth it would be that the poorest $f$ fraction of individuals posses $\mathcal{L}(f)$ fraction of the total wealth of the society. This has been generalized in many other socio-economic contexts (e.g., citations of authors \cite{kolkata}) as well as physical systems (e.g., avalanches in driven disordered systems \cite{diksha}).  

For the present case, the diverging response function $\mathcal{R}$ monotonically grows as the critical point is approached. and the Lorenz function from a point $A$ to another point (closer to criticality) $B$ is given by $\mathcal{L}(f,n,A,B)=\int\limits_{A}^{A+f(B-A)}\mathcal{R}dv/\int\limits_{A}^B\mathcal{R}dv$. Then the Gini index is simply $g(n,A,B)=1-2\int\limits_0^1\mathcal{L}(f,n,A,B)df$. The limiting values of the Gini index, $0$ and $1$, represent the cases of complete equality and extreme inequality respectively.

In the simplest case of a single tuning field (say, temperature for the Ising model), it was shown before \cite{das} that the critical divergence of any response function can be written in terms of the critical Gini index interval $\Delta g=|g_f-g|$, where $g_f$ is when $B$ is exactly the critical point value of the driving field (e.g., the Curie temperature for the Ising model). As we show with the Landau theory argument, at least in the mean field, the formulation is applicable for an arbitrary number of tuning fields, resulting in the following scaling form of any diverging response function
\begin{equation}
    \mathcal{R} \propto |v-v_c|^{-n} \propto |g-g_f|^{-n^*}\propto \Delta g_{n}^{-n^*},
\end{equation}
where, $v_c$ is any point on the critical hyperplane $a_1(u_1,u_2,....,u_r)=0$, and $g_f$ and $n^*$ are sole functions of the (universal) critical exponent $n$. Particularly, $g_f=n/(2-n)$, $n^*=n/(1-n)$ for $0<n<1$; $g_f=1$, $n^*=n/(n-1)$ for $1<n<2$ and $g_f=1$, $n^*=n$ for $n>2$. The cases of $n=1$ and $2$ show logarithmic corrections. Note that the mean field susceptibility and its square face the unique challenge of falling precisely into these two values. However, as we shall shortly see, this is easily circumvented by working with $\chi^3$. In non mean field cases, these situations do not arise. 

Therefore, for any multi parameter system undergoing critical transition, a suitably defined diverging response function (or its sufficiently higher power) can be written in terms of a single parameter viz. the Gini index of the said response function. This implies that the phase diagram is now that of a single parameter, and the value of the parameter at the critical point, $g_f$, is solely a function of the (universal) critical exponent $n$. Indeed, for sufficiently large $n (>2)$, $g_f=1$ for any system, making it the universal signature of critical point for any diverging response function of any model undergoing critical transition (see Fig. \ref{fig_0}). This is a significant simplification, since the scaling behavior as well as the critical point are independent of how the variable (path) $v$ is defined. In situations where tuning parameters are not easily controllable (e.g., the exogeneous and endogeneous forces in ecological transitions \cite{pnas}) or one or more of them are not accurately known (e.g., the porosity of a medium), the critical point and scaling of a response function (e.g., fluctuations in vegetation yields or average avalanche sizes for the above examples) can be predicted, and thereby making it possible to estimate the proximity of a critical transition.  

We will return to the question of a more precise estimation of proximity to critical point or early warning signals. However, now we turn to three specific models where the above framework can be applied. Specifically, we first look into a mean field calculation of the transverse field Ising model and show that the Landau theory predictions and thereby the simplification of the phase surface hold. We also verify this numerically. Then we discuss the same with numerical calculations for site diluted Ising model on the Bethe lattice. Finally, we look at a mean field model of opinion dynamics (essentially undergoing a non-equilibrium active-absorbing phase transition), which also has a tri-critical point and discuss how that might affect the formulation mentioned above.

%Now lets consider the case Transverse Ising model where $u_1=T$ and $u_2=\Gamma$, the order parameter for the continuous transition becomes $m_z\propto\left[\frac{\partial v}{\partial T}\Delta T+\frac{\partial v}{\partial \Gamma}\Delta \Gamma\right]^{\frac{1}{2}}$ and similarly $\chi\propto\left[\frac{\partial v}{\partial T}\Delta T+\frac{\partial v}{\partial \Gamma}\Delta \Gamma\right]^{-1}$ and that's what we see at least in structural form when we solve the mean-filed TIM close to the phase boundary. Which also sugges

%\section{Calculation: Ising model in a transverse field - mean field approach} 

\section{Application to models}
As indicated before, we now discuss the application of the formulation described before in three different models and demonstrate through analytical calculations and numerical simulations that validity of the formulation. These models come from widely different use cases, viz. quantum magnetism, percolation and opinion dynamics. The first two show equilibrium transitions and an energy function (Hamiltonian) can be defined for them. The opinion dynamics model does not have such energy functions and shows a non-equilibrium active-absorbing phase transition. The commonality, of course, is that the framework described in the previous section is applicable to all these models.  
\subsection{Transverse Ising Model in mean field} 
The Hamiltonian for the spin-$\frac{1}{2}$ Ising model in both the longitudinal and transverse field reads
\begin{equation}
    \mathcal{H}= -J\sum_{<ij>} \sigma^{z}_{i}\sigma^{z}_{j} -h\sum_i\sigma^{z}_{i} - \Gamma\sum_{i} \sigma^{x}_{i},
\end{equation}
where, $\sigma^{z}_{i}$ and $\sigma^{x}_{i}$ are the Pauli's spin matrices at the $i^{th}$ lattice site. The first summation accounts for the coupling $J$ between the nearest neighbor spins, the second and third summations account for the Zeeman's energy term in longitudinal magnetic field $h$ and transverse magnetic field $\Gamma$ respectively. The model shows a phase transition from an ordered phase (with $\langle \sigma_z\rangle\ne 0$) to a disordered phase (with $\langle \sigma_z\rangle= 0$), through changing either or both temperature ($T$) and the transverse field ($\Gamma$). The model shows many fascinating properties (see \cite{sach,tim} for details), including their physical manifestations in different materials \cite{tim_book}. 

\begin{figure}
\includegraphics[width=8cm]{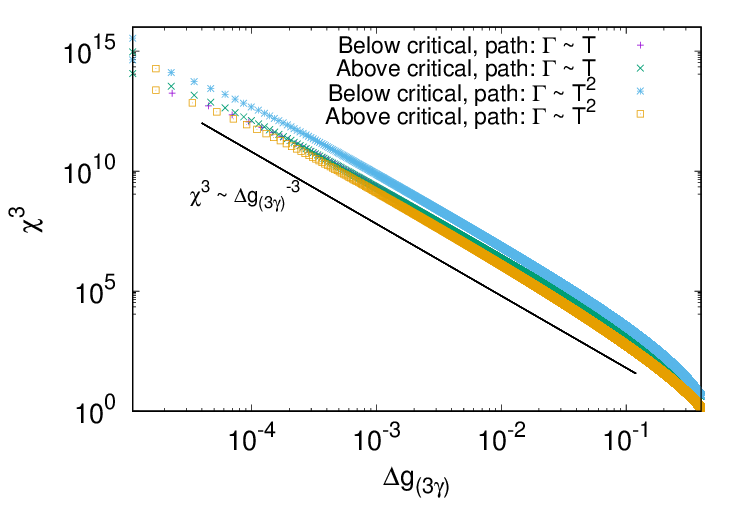}
\caption{The scaling of magnetic susceptibility $\chi$ in terms of the Gini index for two different path for the transverse Ising model in the mean field limit.}
\label{fig_1}
\end{figure}

\begin{figure}
\includegraphics[width=8cm]{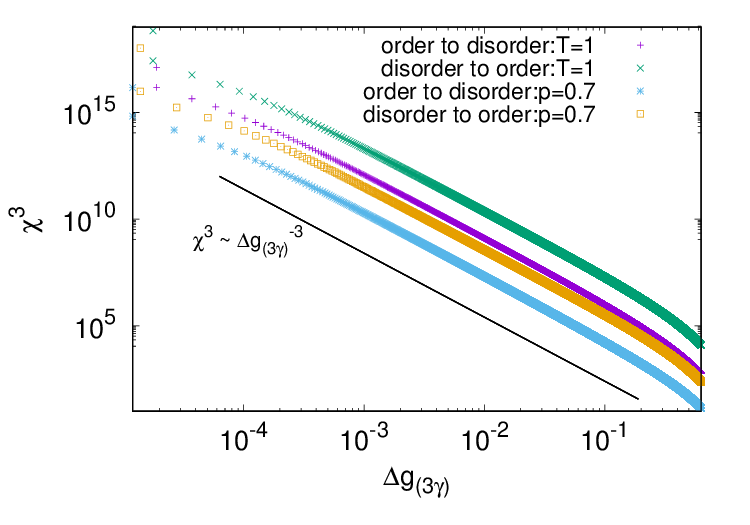}
\caption{The scaling of magnetic susceptibility $\chi^{3}$ in terms of the Gini index from both side of the critical temperature at site concentration $p=0.7$ and from both side of critical site concentration at temperature $T=1$ for site dilution Ising model on bethe lattice. }
\label{fig_2}
\end{figure}

A mean field version of the above Hamiltonian can be written as \cite{mf_tim1,mf_tim2}:    $\mathcal{H}_{MF}=-\sum_i \vec{h_{eff}}.\vec{\sigma_i}$ with an effective field $\vec{h_{eff}}= (qJm_z+h)\hat{z}+\Gamma\hat{x}$, where $m_z=\langle \sigma_z\rangle$ and $h_{eff}\equiv|\Vec{h_{eff}}|=\sqrt{(qJm_z+h)^2+\Gamma^2}$. 
The transverse magnetisation then reads $m_x=\langle \sigma_x\rangle=\tanh{\left(\frac{h_{eff}}{k_BT}\right)} \left(\frac{\Gamma}{h_{eff}}\right)$. The longitudinal magnetisation is
\begin{equation}
    m_z=\left[\tanh{\left(\frac{h_{eff}}{k_BT}\right)}\right] \left(\frac{qJm_z+h}{h_{eff}}\right)
    \label{tr_mag}
\end{equation}
At the phase boundary $m_z=0$, which gives 
\begin{equation}
    \frac{\Gamma_{c}}{qJ}=\tanh{\left(\frac{\Gamma_{c}}{k_BT_c}\right)}.
    \label{PhBound}
\end{equation}
Near to this boundary $m_z$ is small, which implies $\frac{\Gamma}{qJ}\left[1+\left(\frac{q^2J^2m_z^2}{2\Gamma^2}\right)\right]\simeq \tanh{\left(\frac{\Gamma}{k_BT}\right)}$. Then it follows that 
\begin{equation}
    m_z=\sqrt{A(T_c-T)+B(\Gamma_c-\Gamma)},
    \label{ordPara}
\end{equation}
where $A=\frac{2\Gamma_{c}}{qJk_BT_c^2}$ and $B=\frac{2}{q^2J^2}\left(1-\frac{qJ}{k_BT_c}\right)$. 

The magnetic susceptibility $\chi=\lim_{h\to0} {\frac{\partial{{m_z}}}{\partial{h}}}$ has the form
\begin{equation}
%\begin{align}
\begin{split}
    \chi= \left[\frac{1+qJ\chi}{k_BT}\right]\left[\frac{q^2J^2m_z^2}{q^2J^2m_z^2+\Gamma^2}\right]\sech^2{\left[\frac{\sqrt{q^2J^2m_z^2+\Gamma^2}}{k_BT}\right]}\\ 
        + \left[\frac{\Gamma^2(1+qJ\chi)}{(q^2J^2m_z^2+\Gamma^2)^\frac{3}{2}}\right]\tanh{\left[\frac{\sqrt{q^2J^2m_z^2+\Gamma^2}}{k_BT}\right]}.
\end{split}
\label{chi}
%\end{align}
\end{equation}
Near the critical boundary it reads
\begin{equation}
    \chi=\frac{a}{b(\Gamma_c-\Gamma)+c(T_c-T)}
    \label{ChiScaling}
\end{equation}
where $a= -\left[\tanh{\frac{\Gamma_{c}}{k_BT_c}}- \frac{\Gamma_c-\Gamma}{k_BT_c}+\frac{\Gamma_{c}(T_c-T)}{k_BT_c^2}\right]$, $b=\left(1-\frac{qJ}{k_BT_c}\right)$ and $c=\frac{\Gamma_{c}qJ}{k_BT_c^2}$.

Clearly, the scaling forms of Eq. (\ref{ordPara}) and Eq. (\ref{ChiScaling}) match the forms obtained from the Landau expansions (Eq. (\ref{L_op}) and Eq. (\ref{L_chi}) respectively). Specifically, the two tuning fields here are $u_1=T$ and $u_2=\Gamma$. A numerical evaluation of the Gini index (of $\chi^3$ to avoid the logarithmic terms mentioned earlier) and the corresponding scaling of $\chi$ is shown in Fig. \ref{fig_1} along two different paths ($v(T,\Gamma)$): $\Gamma \propto T$ and $\Gamma \propto T^2$ that show universal scaling on both sides of the critical point $g_f=1$. Therefore, the two parameter critical phase surface (Eq. (\ref{PhBound})) is reduced to one critical point.

The above analysis holds for lower dimensions as well, except that the special case of $\Gamma_c=\Gamma_c^0$ for $T=0$, where the model is known to exhibit the critical behavior of the higher dimensional classical Ising model. Above the upper critical dimension, of course, such distinctions are irrelevant.

\subsection {Site diluted Ising model on a Bethe lattice} 
We now consider a site diluted Ising model on a Bethe lattice. The model is a graph that has a central site which connects to $q$ nearest neighbors and then each of the neighbor is connected to $q-1$ further nearest neighbors going outwards (i.e., neglecting the central site) and so on. Deep inside the graph where all sites are equivalent and have $q$ coordination number, is called a Bethe lattice. Then we assign a random variable $c_i$ which equals to unity if there is an Ising spin at site $i$ and zero otherwise. The ensemble average of $\langle c_i\rangle =p$, where $p$ is the probability of any site being occupied by a spin. 
%in in cases beyond the mean field limit, we do the above analysis for the numerical simulations of the site diluted Ising model on a square lattice. This is just the Ising model on the square lattice, which has $p$ fraction of its sites occupied. 
The Hamiltonian without any external force reads
\begin{equation}
    \mathcal{H}= -J\sum_{<ij>}c_ic_j \sigma^{z}_{i}\sigma^{z}_{j},
\end{equation}
%with $c_l$ is $1$ or $0$ depending on whether the $l$-th site is occupied or vacant, respectively.
The model shows critical phase transition with the mean field exponent of the Ising model except at the percolation point $p_c=\frac{1}{q-1}$ i.e., for $T\rightarrow 0$. So a phase boundary exists in the $T-p$ plane. The model is exactly solvable (see e.g. \cite{young}). For our purpose we need a response function that can be the magnetic susceptibility which has the form \cite{young}
\begin{equation}
k_BT\chi=\frac{1}{2}\frac{1+pt}{pt(q-1)-1}
\end{equation}
below the phase boundary and 
\begin{equation}
k_BT\chi=\frac{1+pt}{1-pt(q-1)} 
\end{equation}
above the critical phase surface, where $t=\tanh{\frac{J}{k_BT}}$. We numerically evaluate these two expressions keeping $p$ fixed while varying $T$ and vice-versa, and then measure the Gini index. As the diverging exponent for susceptibility is one, we take the cube of susceptibility, as before, and express its scaling in terms of the Gini index (see Fig. \ref{fig_2})

\begin{figure}
\includegraphics[width=8cm]{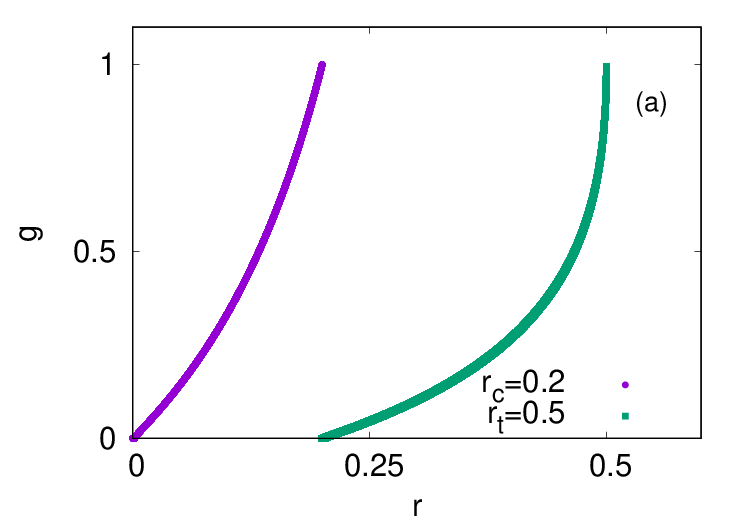}
\includegraphics[width=8cm]{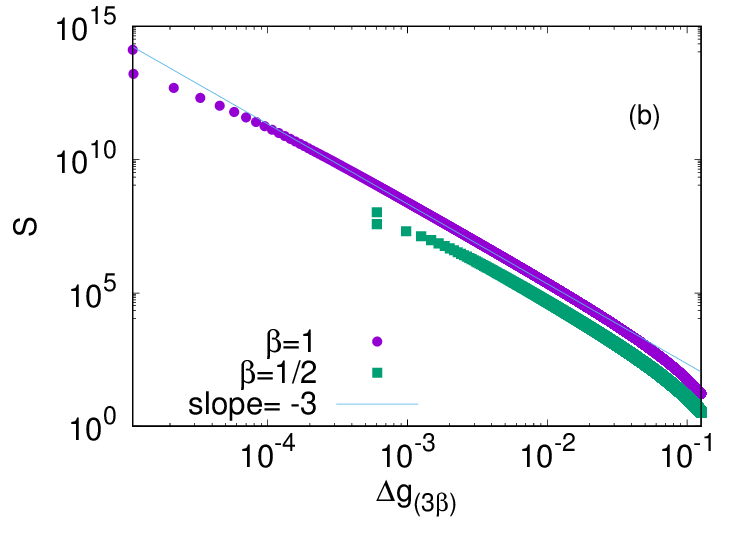}
\caption{(a) The Gini index of the quantity $S=O^{-3}$ is shown as the critical phase surface is approached near $r_c=0.2$ and $r_t=0.5$, in both cases two different linear paths are chosen in the $p-r$ plane from order to disorder. The value of $g$ at the critical point reaches 1, as expected, in both cases. (b) the scaling behavior of $S$ is shown with $\Delta g_{(3\beta)}$, with a scaling exponent $-3$. The subscript indicates the magnitude of the exponent of the response function  $S$.}
\label{fig_4}
\end{figure}

%and  the Ising model being a marginal case according to the Harris criterion (specific heat exponent $\alpha=0$ in the pure case), there are earlier suggestions that the critical exponent values, especially for the cases of low $p$ ($\gtrsim p_c$, the percolation threshold), undergo scaling corrections \cite{kenna,kenna2}. For our purposes, we only stick to rather low dilution limit ($p=0.9$).   

%Fig. \ref{fig_2} shows the scaling of $\chi$ and $\chi^2$, along with the predicted values of the critical exponents showing a reasonable match.  

\subsection{Kinetic exchange opinion model in mean field}
We now turn to a non-equilibrium active-absorbing phase transition in a model of sociophysics \cite{galam} that shows a tri-critical point along with a critical transition (with two tuning parameters).
There have been considerable efforts in modeling interactions among individuals that can eventually lead to formation of consensus in a society \cite{fortunato,soc_ox}. The complications, of course arise from, among other things, quantification of something as abstract as opinion. Also, the interactions among humans stem from complex socio-economic processes and are not simply based on set rules. The first question is addressable rather easily, given there exist several situations where opinions are effectively discrete --- elections in two-party systems, voting in favor or against an issue (e.g., Brexit \cite{brexit}) etc. In such cases, an Ising spin like variable, having values $\pm 1$ and 0, representing two choices and neutrality, can be invoked. The second issue is circumvented with the argument that like in a critical transition where an emergent correlation (length)  suppresses many details of the system and the critical behavior is then independent of such details, in these systems also the nuances of individual interactions are suppressed in favor of a simple set of rules that can reproduce the essential characteristics of opinion dynamics (see e.g., \cite{santh} for a recent effort towards this). 

Here we look at a kinetic exchange model for opinion formation \cite{mfke}. Our purpose is to take this as an example of a non-equilibrium active-absorbing type transition and the applicability of the framework described above in this context. In this variant of the model, two or more agents interact at a time and their opinion values can get modified following this exchange. The changes depend on the conviction of the interacting agents in their earlier opinions and a stochastic term determining the nature of interaction. 

In the present context, we denote the individual opinion of $i$-th agent at time $t$ by $o_i(t)$, which can take discrete values $\pm 1$ and $0$. When two agents interact, their opinion values change following (we only write the $i$-th agent's opinion, similar equation holds for $j$-th agent)
\begin{equation}
    o_i(t+1)=\lambda o_i(t)+\lambda\epsilon o_j(t),
    \label{two-body}
\end{equation}
where $\lambda$ is the conviction parameter (taken same for all to maintain simplicity) that takes the value $\lambda=1$ with probability $p$ and $\lambda=0$ with probability $1-p$; $\epsilon$ takes the value $0$ or $1$ with equal probability (1/2) randomly for each interaction. If following the equations takes the value of $o_i(t+1)$ to be greater than $1$ (or less than $-1$), then it is limited at $1$ (or $-1$), representing the two extreme choices. 
We then also have a three-body exchange, which is usually discarded in kinetic theories of gas molecules from which these models draw their original inspirations. But in the case of opinion exchanges, these are very much possible. A three-agent exchange follow:
\begin{equation}
    o_i(t+1)=\lambda o_i(t)+\lambda\epsilon\theta_{jk}(t),
    \label{three-body}
\end{equation}
where, 
\begin{equation}
\theta_{jk}(t)= \begin{cases} o_j(t) &\text{if  $o_j(t)=o_k(t)$}  \\
 0 &\text {otherwise.} \nonumber
\end{cases}
\end{equation}

During the dynamics of the model, at each time step, either a three-agent interaction (following Eq. (\ref{three-body})) occurs with probability $r$, or a two-body interaction occurs (following Eq. (\ref{two-body})) with probability $1-r$. The agents are chosen randomly from the total number of $N$ agents, as this is a mean field model. The limits in the individual opinion values, as mentioned above, always apply. $N$ such updates constitute one time step. Then, depending on the two parameters $p$ and $q$, the model undergoes an active (finite fraction of opinion values are non-zero) to absorbing (all opinion values are $0$) phase transition. It is straightforward to show \cite{mfke} that the equation of the continuous transition line is $p_c=2/(3-r_c)$ that terminates at the point $p_t=1/2, q_t=4/5$, which is a tri-critical point. For $r>r_t=1/2$, there is a discontinuous transition along the line $p_c=8r_c/(1+8r_c)$.  

The order parameter for this transition, which is the sum of the fraction of individuals with opinion value $+1$ and that with opinion value $-1$ is given by
\begin{equation}
  O=\left(\frac{2r-1}{2r}+\frac{\sqrt{p^2/4-2pr(1-p)}}{pr}\right).
\label{mf-op}
\end{equation}
Near a critical transition, this can be written as
\begin{equation}
    O\sim A_1\Delta_1+A_2\Delta_2,
\end{equation}
where, $\Delta_1=p-p_c$, $\Delta_2=r_c-r$ and $A_1=\frac{50p_c-8r_c-32}{12p_c-8}-r_c\frac{25p_c^2-40p_c+16}{18p_c^2-12p_c+8}$,   $A_2=\frac{8p_c-8p_c^2}{12p_c-8}+p_c\frac{25p_c^2-40p_c+16}{18p_c^2-12p_c+8}-\frac{1}{2r_c^2}$ are constants that are functions of $p_c,r_c$. Similarly, near the tri-critical point ($p_t=4/5, r_t=1/2$), it reads
\begin{equation}
    O\sim \left(A_1^{\prime}\Delta_1^{\prime}+A_2^{\prime}\Delta_2^{\prime}\right)^{1/2},
\end{equation}
where, $\Delta_1^{\prime}=p-p_t$, $\Delta_2^{\prime}=r_t-r$ and $A_1^{\prime}=\frac{50p_t-8r_t-32}{36p_t^2-24p_t+16}$, $A_2^{\prime}=\frac{8p_t-8p_t^2}{36p_t^2-24p_t+16}$ are constants which are function of $p_t, r_t$.
Both of these are similar in form as that shown in Eq. (\ref{L_op}), although a Landau-like approach is not possible here.

For this model, there is no obvious definition of susceptibility either. However, as discussed earlier, we simply need a response function that diverges as the critical transition (or the tri-critical point) is approached. For the case $r\le 1/2$, we just take the inverse of the cube of the order parameter, denoted here as $S=O^{-3}$ for our subsequent analysis, which will behave as $S\sim \left(A_1\Delta_1+A_2\Delta_2\right)^{-3}$ and $S\sim \left(A_1^{\prime}\Delta_1^{\prime}+B_1^{\prime}\Delta_2^{\prime}\right)^{-3/2}$ for the critical transition and the tri-critical point respectively. Note that if we had taken $O^{-1}$ or $O^{-2}$, for the critical transition we would run into a logarithmic form for the Gini index, as indicated above.

Then, for both of these cases (critical transition and the tri-critical point), we expect $g_f=1$, and coincidentally, we also expect the rescaled critical exponents for $S$ to be the same: for the critical transition $n^*=n=3$ (since $n=3$) and for the tri-critical point $n^*=n/(n-1)=3$ (since $n=3/2$ here).
Numerical evaluation of the Gini index for $S$ near the critical line (near a chosen point $p_c= , r_c= $)
show $g_f=1$ and similarly that near the tri-critical point $p_t=4/5, r_t=1/2$ shows $g_f=1$ (see Fig. \ref{fig_4}(a)). The scaling of $S$, in terms of $g_f-g$ also verify $n^*=3$ for both of these cases (see Fig. \ref{fig_4}(b)). If we had taken an even higher power of inverse order parameter, the rescaled exponent values would differ for the critical transition and the tri-critical point, but $g_f$ value ($1$) would be the same.

These results show that the method is valid for non-equilibrium active-absorbing transitions as well. It also shows the possibility extension of the framework to tri-critical points. Of course, with our choice of the response function, it is not possible to implement this procedure in crossing the transition surface from disordered to ordered state, as the order parameter is identically zero in the absorbing phase. But that can also be achieved by some other choice of a diverging response function, for instance the relaxation time.

\section{Early Warning Signals (EWS)} 
Finally, an important question in systems with transitions having potentially catastrophic consequences is the estimation of an imminent transition point. Many such Early Warning Systems have been studied \cite{ew1,ew2,ew3}. Here, we look at this question with the help of another inequality measure, called the Kolkata index ($k$) \cite{kolkata}, defined as the $1-k$ fraction of the largest values posses $k$ fraction of all values. In socio-economic contexts, it is a generalisation of the Pareto's 80-20 law. It can be calculated from the Lorenz function by evaluating the fixed point $1-k=\mathcal{L}$. 

\begin{figure}
\includegraphics[width=8cm]{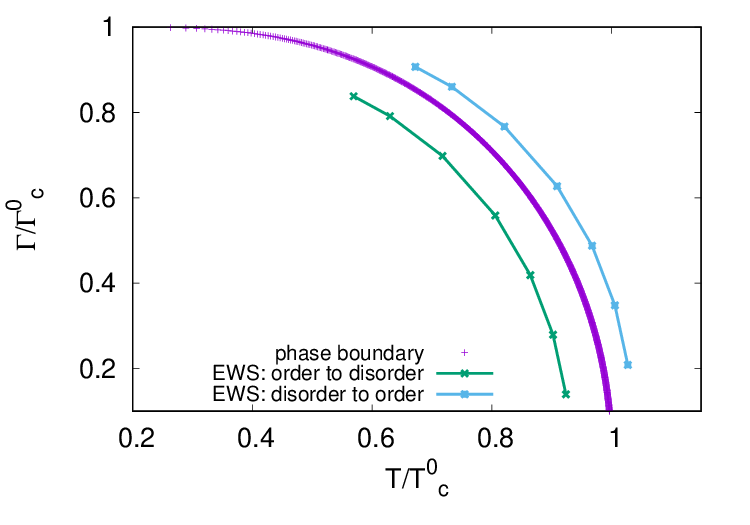}
\caption{The phase boundary of the transverse Ising model and its Early Warning Signal boundaries on both sides, drawn using the crossing points of $g$ and $k$.}
\label{fig_3}
\end{figure}

In conjunction with the Gini index, the Kolkata index plays an important role in designing Early Warning Signals for imminent transition. Particularly, it has been shown that for single parameter models, $g$ and $k$ cross each other at a near-universal value ($\sim 0.87$) \cite{manna,das} (see also \cite{lomov}).  Other than theoretical models, this framework for Early Warning Signals has been shown to work in experimental data of fracture of disordered solids \cite{diksha1}. 

This formalism also applies to multi parameter systems. Specifically, we show in Fig. \ref{fig_3}, the phase boundary for the transverse field Ising model (Eq. \ref{PhBound}) and the Early Warning Signal boundaries, drawn from the crossing points of $g$ and $k$ (for a common starting points which is (0,0) for order to disorder EWS boundary and (1,1.5) for disorder to order EWS boundary). 

The EWS boundaries are always parallel to the critical phase surface, signaling that when it is reduced to a single parameter, the EWS boundary is simply a point that can be easily estimated from the signals of a suitable response function.   

\section{Discussions and Conclusions} 
The critical transition point of a system, unlike the critical exponent values, are dependent on specific details of the system. The critical phase (hyper) surface can, therefore, be formed with two or more such parameters e.g., temperature and transverse field for the transverse Ising model, temperature and site dilution for the dilute Ising model and so on. We have shown here that all such cases of critical transition surface, can be reduced to a single critical point that is the Gini index value of a suitable response function of the system at the critical surface, which is a universal number at least for a given universality class. The critical scaling can also be obtained in terms of the Gini index, by crossing the critical surfaces along different paths and from either sides of the transitions. Particularly, we have demonstrated this through analytical calculations and numerical evaluations in the cases of two models having equilibrium transitions -- the transverse field Ising model in the mean field limit (see Fig. \ref{fig_1}) and site diluted Ising model on Bethe lattice (see Fig. \ref{fig_2}), and in one model with non-equilibrium transition -- a discretized version of the kinetic exchange opinion model (see Fig. \ref{fig_4}). The last case includes a tri-critical point, where the method is also shown to work, coincidentally with the same Gini index value (and same critical scaling) with that near the critical transition surface. 

This is a drastic simplification in construction of an otherwise potentially complex phase diagram of an arbitrary system. More importantly, in systems where it is not straightforward to control one or more of the parameters influencing an approach to a critical transition (e.g., plantation yield \cite{pnas}, polar ice coverage \cite{polar}, fracture of porous solids \cite{diksha1}), one can still obtain the critical scaling behavior in terms of the Gini index. Also, in cases where such transitions come with potentially catastrophic consequences, an Early Warning Signal can be obtained, regardless of the path to approach the said transition (see Fig.  \ref{fig_3}, see also ref. \cite{diksha1}).  

In conclusion, the multi-parameter critical phase surface of any model, with a second order transition across it, can be reduced to a single parameter boundary (point) when represented in terms of the Gini index of a suitably defined diverging response function of the system. We demonstrated its working analytically and numerically for the mean field transverse Ising model, site diluted Ising model on a Bethe lattice and kinetic exchange opinion dynamics models showing non-equilibrium active-absorbing transition, including a tri-critical point. This simplification paves the way for an unambiguous Early Warning Signal for complex many parameter systems approaching a critical transition. 

\section*{Acknowledgements} The authors are grateful to Bikas K. Chakrabarti for fruitful comments on the manuscript. The simulations were performed in the HPCC Surya at SRM University - AP.

\end{document}